\documentclass[twocolumn,showpacs,aps,prd,superscriptaddress]{revtex4}

\usepackage{graphicx}
\usepackage{dcolumn}
\usepackage{amsmath}
\usepackage{epsfig}

\RequirePackage{xspace}

\usepackage{relsize}
\def\babar{\mbox{\slshape B\kern-0.1em{\smaller A}\kern-0.1em
    B\kern-0.1em{\smaller A\kern-0.2em R}}}

\def\epem       {\ensuremath{e^+e^-}\xspace}

\def\qqbar {\ensuremath{q\overline q}\xspace}

\def\piz   {\ensuremath{\pi^0}\xspace}

\def\pip   {\ensuremath{\pi^+}\xspace}
\def\pim   {\ensuremath{\pi^-}\xspace}
\def\pipi  {\ensuremath{\pi^+\pi^-}\xspace}

\def\Kbar  {\kern 0.2em\overline{\kern -0.2em K}{}\xspace}

\def\Kz    {\ensuremath{K^0}\xspace}
\def\Kzb   {\ensuremath{\Kbar^0}\xspace}
\def\KzKzb {\ensuremath{\Kz \kern -0.16em \Kzb}\xspace}
\def\Kp    {\ensuremath{K^+}\xspace}
\def\Km    {\ensuremath{K^-}\xspace}

\def\KpKm  {\ensuremath{\Kp \kern -0.16em \Km}\xspace}
\def\KS    {\ensuremath{K^0_{\scriptscriptstyle S}}\xspace}

\def\Kstarm  {\ensuremath{K^{*-}}\xspace}

\def\Dbar    {\kern 0.2em\overline{\kern -0.2em D}{}\xspace}

\def\Dz      {\ensuremath{D^0}\xspace}
\def\Dzb     {\ensuremath{\Dbar^0}\xspace}
\def\DzDzb   {\ensuremath{\Dz {\kern -0.16em \Dzb}}\xspace}
\def\Dp      {\ensuremath{D^+}\xspace}
\def\Dm      {\ensuremath{D^-}\xspace}

\def\DpDm    {\ensuremath{\Dp {\kern -0.16em \Dm}}\xspace}

\def\B       {\ensuremath{B}\xspace}
\def\Bbar    {\kern 0.18em\overline{\kern -0.18em B}{}\xspace}

\def\BB      {\ensuremath{B\Bbar}\xspace} 
\def\Bz      {\ensuremath{B^0}\xspace}
\def\Bzb     {\ensuremath{\Bbar^0}\xspace}
\def\BzBzb   {\ensuremath{\Bz {\kern -0.16em \Bzb}}\xspace}
\def\Bu      {\ensuremath{B^+}\xspace}
\def\Bub     {\ensuremath{B^-}\xspace}

\def\Bm      {\ensuremath{\Bub}\xspace}

\def\BpBm    {\ensuremath{\Bu {\kern -0.16em \Bub}}\xspace}

\def\BorBbar    {\kern 0.18em\optbar{\kern -0.18em B}{}\xspace}
\def\DorDbar    {\kern 0.18em\optbar{\kern -0.18em D}{}\xspace}
\def\KorKbar    {\kern 0.18em\optbar{\kern -0.18em K}{}\xspace}

\mathchardef\Upsilon="7107
\def\Y#1S{\ensuremath{\Upsilon{(#1S)}}\xspace}% no space before {...}!

\def\FourS {\Y4S}

\mathchardef\Deltares="7101
\mathchardef\Xi="7104
\mathchardef\Lambda="7103
\mathchardef\Sigma="7106
\mathchardef\Omega="710A

\def\Deltabar{\kern 0.25em\overline{\kern -0.25em \Deltares}{}\xspace}
\def\Lbar{\kern 0.2em\overline{\kern -0.2em\Lambda\kern 0.05em}\kern-0.05em{}\xspace}
\def\Sigbar{\kern 0.2em\overline{\kern -0.2em \Sigma}{}\xspace}
\def\Xibar{\kern 0.2em\overline{\kern -0.2em \Xi}{}\xspace}
\def\Obar{\kern 0.2em\overline{\kern -0.2em \Omega}{}\xspace}
\def\Nbar{\kern 0.2em\overline{\kern -0.2em N}{}\xspace}
\def\Xb{\kern 0.2em\overline{\kern -0.2em X}{}\xspace}

\def\BR         {{\ensuremath{\cal B}\xspace}}

\def\mes        {\mbox{$m_{\rm ES}$}\xspace}

\def\DeltaE     {\mbox{$\Delta E$}\xspace}

\newcommand{\tev}{\ensuremath{\mathrm{\,Te\kern -0.1em V}}\xspace}
\newcommand{\gev}{\ensuremath{\mathrm{\,Ge\kern -0.1em V}}\xspace}
\newcommand{\mev}{\ensuremath{\mathrm{\,Me\kern -0.1em V}}\xspace}
\newcommand{\kev}{\ensuremath{\mathrm{\,ke\kern -0.1em V}}\xspace}
\newcommand{\ev}{\ensuremath{\mathrm{\,e\kern -0.1em V}}\xspace}
\newcommand{\gevc}{\ensuremath{{\mathrm{\,Ge\kern -0.1em V\!/}c}}\xspace}
\newcommand{\mevc}{\ensuremath{{\mathrm{\,Me\kern -0.1em V\!/}c}}\xspace}
\newcommand{\gevcc}{\ensuremath{{\mathrm{\,Ge\kern -0.1em V\!/}c^2}}\xspace}
\newcommand{\mevcc}{\ensuremath{{\mathrm{\,Me\kern -0.1em V\!/}c^2}}\xspace}
\def\cm   {\ensuremath{\rm \,cm}\xspace}

\def\mm   {\ensuremath{\rm \,mm}\xspace}

\def\invfb   {\ensuremath{\mbox{\,fb}^{-1}}\xspace}

\def\mus  {\ensuremath{\rm \,\mus}\xspace}

\def\mus        {\ensuremath{\,\mu{\rm s}}\xspace}    %% microsecond
\def\to                 {\ensuremath{\rightarrow}\xspace}

\def\pep2{PEP-II}

\newcommand{\dedx}{\ensuremath{\mathrm{d}\hspace{-0.1em}E/\mathrm{d}x}\xspace}

\def\gsim{{~\raise.15em\hbox{$>$}\kern-.85em
          \lower.35em\hbox{$\sim$}~}\xspace}
\def\lsim{{~\raise.15em\hbox{$<$}\kern-.85em
          \lower.35em\hbox{$\sim$}~}\xspace}

\def\CP                {\ensuremath{C\!P}\xspace}
\xspace

\def\jetset74   {\mbox{\tt Jetset \hspace{-0.5em}7.\hspace{-0.2em}4}\xspace}

\newcommand{\BRBmtoDzKstarm}{\ensuremath{\BR(\BmtoDzKstarm)}\xspace}
\newcommand{\BmtoDzKstarm}{\ensuremath{\Bm \to \Dz\Kstarm}\xspace}
\newcommand{\BmtoDzpim}{\ensuremath{\Bm \to \Dz\pim}\xspace}
\newcommand{\BmtoDzKm}{\ensuremath{\Bm \to \Dz\Km}\xspace}
\newcommand{\DztoKPi}{\ensuremath{\Dz \to \Km\pip}\xspace}
\newcommand{\DztoKPiPiz}{\ensuremath{\Dz \to \Km\pip\piz}\xspace}
\newcommand{\DztoKThreePi}{\ensuremath{\Dz \to \Km\pip\pim\pip}\xspace}
\newcommand{\KPi}{\ensuremath{\Km\pip}\xspace}
\newcommand{\KPiPiz}{\ensuremath{\Km\pip\piz}\xspace}
\newcommand{\KThreePi}{\ensuremath{\Km\pip\pim\pip}\xspace}

\newcommand{\CosThT}{\ensuremath{|\cos \theta_T|}\xspace}
\newcommand{\CosThH}{\ensuremath{\cos \theta_H}\xspace}

\newcommand{\BABARPubYear}    {03}
\newcommand{\BABARPubNumber}  {035}

\newcommand{\SLACPubNumber} {10285}

\def\figurebox#1#2#3{%
    \def\arg{#3}%
    \ifx\arg\empty
    {\hfill\vbox{\hsize#2\hrule\hbox to #2{\vrule\hfill\vbox to #1{\hsize#2\vfill}\vrule}\hrule}\hfill}%
    \else
    {\hfill\epsfbox{#3}\hfill}%
    \fi}

\begin{document}

\preprint{\babar-PUB-\BABARPubYear/\BABARPubNumber} 
\preprint{SLAC-PUB-\SLACPubNumber} 

\begin{flushleft}
\babar-PUB-\BABARPubYear/\BABARPubNumber\\
SLAC-PUB-\SLACPubNumber\\
%#hep-ex/\LANLNumber\\[20mm]
\end{flushleft}

\title{
%\vskip 10mm
{\large \bf
Measurement of the Branching Fraction for {\boldmath{\BmtoDzKstarm}}} 
}
%% author list as of 01-Sep-2003 (603 authors)
%
\author{B.~Aubert}
\author{R.~Barate}
\author{D.~Boutigny}
\author{F.~Couderc}
\author{J.-M.~Gaillard}
\author{A.~Hicheur}
\author{Y.~Karyotakis}
\author{J.~P.~Lees}
\author{P.~Robbe}
\author{V.~Tisserand}
\author{A.~Zghiche}
\affiliation{Laboratoire de Physique des Particules, F-74941 Annecy-le-Vieux, France }
% ...
\collaboration{The \babar\ Collaboration}
\noaffiliation

%\date{\today}% It is always \today, today, but you may specify any date with \date.
\date{\today}

\begin{abstract}
We present a measurement of the branching fraction for the decay \BmtoDzKstarm using
a sample of approximately 86 million \BB pairs collected by the 
\babar\ detector
from \epem collisions near the \FourS resonance. 
The \Dz is
detected through its decays to \KPi, \KPiPiz and \KThreePi, and the
\Kstarm through its decay to \KS\pim. 
We measure the branching fraction to be
\BRBmtoDzKstarm= (6.3$\pm$0.7(stat.)$\pm$0.5(syst.))$\times 10^{-4}$.
\end{abstract}

% PACS, the Physics and Astronomy Classification Scheme.

\maketitle

A comprehensive test of \CP\ violation within the Standard Model requires 
precision measurements of
the three sides and three angles of the Unitarity Triangle, which
are combinations of various Cabibbo-Kobayashi-Maskawa (CKM) matrix 
elements~\cite{CKM}. 
The measurement of the angle $\gamma$ of the Unitarity Triangle 
is challenging and requires larger samples of
\B mesons than are currently available. A precise determination of $\gamma$ 
at the \B factories is likely to use many different decay modes.
Decays of the form $\B\to D^{(*)} K^{(*)}$ can 
provide a theoretically clean determination of $\gamma$~\cite{gamma}.
For some of the proposed methods, there are 
distinct advantages  to using the K* modes~\cite{gammakstar}.
In this paper, we measure the branching fraction for one of these decays, 
\BmtoDzKstarm~\cite{chargeconj}, which was first observed by the CLEO experiment~\cite{CLEO_BF}.
If the \Dz is reconstructed in its decay to \CP\ eigenstates, the $b\to c \bar{u} s$
and $b\to u \bar{c} s$ quark transitions interfere, giving access to 
the phase $\gamma$ through the measurement of direct \CP\ violation asymmetries.
However, the branching fractions for \Dz decays to \CP\ eigenstates are 
only of the order of 1\%, too small for the size of the available data sample. 
Therefore, for this analysis, we use
decay modes of the \Dz and \Kstarm that have clear experimental signatures and
sufficiently high branching fractions. This measurement provides an important
first step towards establishing the feasibility of using the decay \BmtoDzKstarm
for a future determination of $\gamma$.

We present here a measurement of the branching
fraction for the decay \BmtoDzKstarm
using data collected 
with the \babar\ detector
at the \pep2\ \epem storage ring.
The data correspond to an integrated luminosity of 81.5\invfb
taken at center-of-mass energies close to the \FourS resonance, giving  
a sample of approximately 86 million \BB pairs. 
We reconstruct \Dz candidates through the decays  
\DztoKPi, \DztoKPiPiz and \DztoKThreePi. \Kstarm candidates are identified 
through the decay \Kstarm\to\KS\pim, with the \KS decaying to a pair of charged
pions. 

A detailed description of the \babar\ detector can be found elsewhere~\cite{ref:babar}. 
Only detector components relevant to this analysis are described here.
Charged-particle trajectories are measured by 
a five-layer double-sided silicon vertex tracker (SVT) and a 40-layer drift chamber (DCH), 
operating in the field of a 1.5-T solenoid. 
Charged-particle identification is achieved by combining 
measurements of ionization energy loss (\dedx) 
in the DCH and SVT
with information from a detector of internally reflected Cherenkov light (DIRC).   
Photons are detected in a CsI(Tl) electromagnetic calorimeter (EMC). 

We set the event-selection criteria 
to minimize the statistical error on the branching fraction, using
simulations of the signal and background. 
In general, charged tracks are
required to have at least 12 DCH hits and a minimum transverse momentum of 0.1\gev, and
to originate from the interaction point, within 10\cm along the beam direction and 1.5\cm 
in the transverse plane. We use less restrictive selection criteria  
for tracks used to reconstruct
$\KS\to\pipi$ candidates, to allow for displaced \KS decay vertices.
Photon candidates are identified in the EMC as deposits of energy isolated from
charged tracks. They are required to 
have a minimum energy of 30\mev and a shower shape consistent 
with that of a photon. 

We use  pairs of photons to reconstruct \piz candidates, which are required
to have an invariant mass between 125 and 144\mev. We reconstruct
\KS candidates from pairs of oppositely charged tracks fitted to
a common vertex. They are required to have an
invariant mass within 8\mev of the \KS mass~\cite{ref:pdg}. 

To reconstruct \Kstarm candidates, we combine
\KS candidates with charged tracks. We require  
the \Kstarm candidate to have an invariant mass within 75\mev of 892\mev.
In addition, the \KS vertex is required to be displaced by at least 3\mm from the \Kstarm
vertex.    

We reconstruct \Dz candidates from the appropriate combination of tracks and \piz candidates.
The \Km tracks must satisfy kaon identification criteria resulting
in an efficiency of 80\%--95\% depending on the
momentum. The probability of a pion to be misidentified as a kaon is less than 5\%.
We require the momenta of the \Km candidates to be greater than 250\mev and their
polar angle (relative to magnetic-field axis) to be in the interval $0.25<\theta<2.55$ rad to restrict
them to a fiducial region where the kaon identification performance
can be determined with small uncertainty.
The tracks from the \Dz are fitted to a common vertex and we accept candidates if they 
have an invariant mass within $18$ ($14$) \mev of the \Dz mass
for the \KPi(\KThreePi) decay. For the \KPiPiz decay, we use an asymmetric mass 
requirement $-29 < (m-1865\mev) <+24\mev$, reflecting the distribution of the energy of the photons from the \piz decay.
It is known that the decay \DztoKPiPiz occurs predominantly through 
an intermediate state (\Kstarm(892) or $\rho^+(770))$. Hence,
to reduce the combinatorial background in the \KPiPiz decay, we select events in
the enhanced
regions of the Dalitz plot, using amplitudes and phases determined by the CLEO
experiment~\cite{DalitzWeight}.

In reconstructing the decay chain, the measured momentum vector of each intermediate particle
is determined by refitting the momenta
of its decay products, constraining the mass to the nominal mass of the particle and requiring 
the decay products to originate from a common point.
For the \Kstarm resonance only a geometrical constraint is used in this kinematic fit.
Finally, to reconstruct \Bm decays, \Dz candidates are combined with \Kstarm candidates. 

The dominant 
background is from $\epem\rightarrow\qqbar$ production.
We suppress this background using requirements on the event topology and kinematics, 
and through the use of a Fisher discriminant.
The ratio of the second and zeroth Fox-Wolfram moments~\cite{ref:FoxWol}, which is a 
measure of the event sphericity and is close to zero for approximately spherical events, 
is required to be less than 0.5. 
The absolute value of the
cosine of the angle between the thrust axis of the \B candidate
and the thrust axis of the rest of the event,~\CosThT,
is peaked at one for continuum events and is approximately flat for \B decays.
We require $\CosThT<0.8$ for \KPi and \KPiPiz decays
and $\CosThT<0.75$ for \KThreePi decays. 
The Fisher discriminant is
built from the momentum of all particles in the event (excluding those
used to form the \B candidate) and the angle between this
momentum and the thrust axis of the reconstructed \Bm, both in the center-of-mass frame~\cite{babarfisher}.
The \Kstarm helicity angle, $\theta_H$, 
defined as the angle between the $\pi^-$ from the \Kstarm decay and the \Bm
flight direction in the rest frame of the \Kstarm,
follows a $\cos^2 \theta_H$ distribution for signal events and is
approximately flat for continuum events. To further reject 
continuum background in the \KThreePi channel, we require
$|\CosThH|>0.4$. 

The selection criteria just described reject
all but approximately $0.001\%$ of the background, 
while retaining between 4\% and 13\% of the signal, depending on the \Dz mode.
In the case of events with more than one \Bm candidate (5--17\%, depending
on the \Dz mode), we choose the best candidate on the basis of the $\chi^2$
formed from the differences of the measured and true \Bm, \Dz, and \KS masses, scaled by the
mass resolutions.
Studies of simulated signal events have determined that
the algorithm does not introduce a bias 
and chooses the correct \Bm candidate in approximately 80\% of the events with
multiple candidates.

Finally, we identify \B-meson decays kinematically using two nearly independent variables:
the energy-substituted \B mass $\mes = \sqrt{(s/2+{\bf p_0}\cdot{\bf p_B})^2/E_0^2-p_B^2}$,
where the subscripts 0 and \B refer to the \epem system and the \B candidate respectively,
$s$ is the square of the center-of-mass energy,
and energies ($E$) and momentum vectors (${\bf p}$) are computed in the laboratory frame; and 
$\DeltaE=E_B^*-\sqrt{s}/2$, where $E_B^*$ is the \B candidate energy in the center-of-mass frame.
We select \Bm candidates with $|\DeltaE|<25\mev$, which corresponds
to approximately $\pm 2.2 \sigma$ (where the resolution $\sigma$ 
is found to be independent of the \Dz decay mode).
In addition, the signal events are expected to have values of \mes close to the \Bm
mass. 

We determine the signal yield of \BmtoDzKstarm events by performing an unbinned maximum likelihood fit
to the \mes distribution of the selected candidates for the signal region in \DeltaE.
The signal distribution is parameterized
as a Gaussian function and the combinatorial background as a threshold function~\cite{ARGUS}.
All parameters except the endpoint of the threshold function are unconstrained in the fit.

The signal yield determined from the fit potentially includes backgrounds from other \BB decays
that also peak in \mes. To investigate this, we have studied a simulated sample of  
generic \BB decays
and also high statistics simulated samples of other $B\to D^{(*)}K^{(*)}$ decays. 
The simulation indicates no enhancement in the signal region from this background.
Therefore, we assume that the
peaking background is negligible and the uncertainty in its determination
from the studies of various simulated event samples is included as a systematic error.
We have also verified that use of the \Bm mass and error
in the $\chi^2$ calculation for the choice of the best \Bm candidate 
does not affect the smooth shape of the background in \mes.

\begin{figure}[htb]
\begin{center}
\epsfig{file=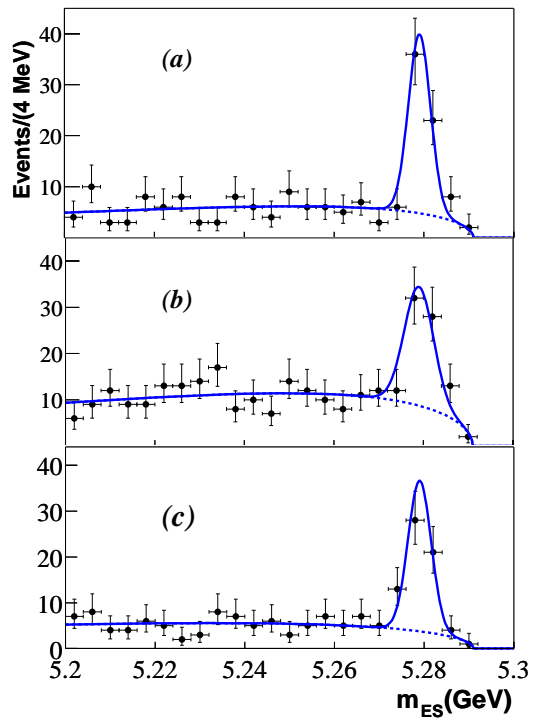,width=8cm}
\caption[]{The \mes distributions of \BmtoDzKstarm candidates: (a) \DztoKPi, (b) \DztoKPiPiz, 
and (c) \DztoKThreePi. The solid lines show the fit used to extract the signal
yields, with the distribution parameterized as a Gaussian plus a threshold function 
as described in the text. The dashed line indicates the combinatorial background component.}
\label{fig:data}
\end{center}
\end{figure}

Figure~\ref{fig:data} shows the \mes distribution for the three different
\Dz decay modes with the fit function superimposed. 
A clear signal is seen in all cases.
The signal yield 
and the size of the combinatorial background in the signal region, i.e., 
for events with $\mes>5.270\gev$,  are detailed in Table~\ref{tab:yield}. 
We observe 
a total of $161\pm17$ \BmtoDzKstarm events.
We have studied the \CosThH distribution for the selected 
candidates and determined that the data are consistent with pure \BmtoDzKstarm
decay.

\begin{table}[tbh]
\caption{Signal yield, number of background events, 
and efficiency for the three \Dz decay modes in the signal region ($\mes>5.270\gev$).
Yields are extracted from the fits to the \mes distribution from data
(errors are statistical only). Efficiencies are computed from 
simulated events.
}
\center
\begin{tabular}{ l c c c }
\hline\hline 
    & \KPi & \KPiPiz & \KThreePi\\
\hline
Signal Yield & 56.2 $\pm$ 9.4 & 51.7 $\pm$ 11.0 & 52.6 $\pm$ 8.7 \\
Background & 19.5 $\pm$ 4.3 & 37.7 $\pm$ 6.2 & 16.4 $\pm$ 3.6 \\
\hline
Efficiency (\%) & 12.8 & 3.5 & 4.0 \\
\hline\hline
\end{tabular}
\label{tab:yield}
\end{table}

We determine 
the selection efficiency 
for each sample of \BmtoDzKstarm events from samples of simulated signal events.
We apply small corrections 
determined from data to the efficiency calculation
to account for the overestimation of the tracking
and particle-identification performance, and 
of the \piz and \KS reconstruction efficiencies in the Monte Carlo simulation.
The product of these efficiency corrections is about 0.9. 

To quantify the ability of the simulation to model the variables used in the event selection,
we use a sample of \BmtoDzpim events from data and Monte Carlo simulation. 
This sample is kinematically 
similar to the \BmtoDzKstarm decay. We select \BmtoDzpim events in 
the same way as the \BmtoDzKstarm sample, with the additional requirement that
the \pim fails loose
kaon identification criteria, to remove \BmtoDzKm events. 
Approximately 3000 \BmtoDzpim candidates in each \Dz 
decay mode are selected from the data. 
The purity of the sample is 94\% for the \KPiPiz decay and
98\% for the \KPi and \KThreePi decays. 
We use this sample to determine correction factors for the efficiencies
for the \BmtoDzKstarm selection.
The obtained correction factors vary from about 0.95 for the \KPi and \KThreePi 
decays to 0.85 for the \KPiPiz decay.
We include the statistical precision of these corrections in the systematic error
of the branching fraction.
The selection efficiency after all corrections are
reported in Table~\ref{tab:yield}.

We determine the branching fraction 
separately for each of the \Dz decay modes
from:
\begin{eqnarray*}
\BRBmtoDzKstarm=\frac{N}{\epsilon \cdot N_{\BB} \cdot \BR_{\Dz}
\cdot \BR_{\Kstarm} \cdot  \BR_{\KS} \cdot \BR_{\piz} }
\end{eqnarray*}
for a signal yield of $N$ events, an efficiency $\epsilon$ 
and a sample containing $N_{\BB}$ pairs of \B mesons.
$\BR_{\Dz}$, $\BR_{\Kstarm}$, $\BR_{\KS}$ and $\BR_{\piz}$, 
the branching fractions for the \Dz, \Kstarm, \KS and \piz respectively to the relevant final states,  
are obtained from Ref.~\cite{ref:pdg} (${\cal B}_{\pi^0}$ in equation is 
only relevant for the \KPiPiz mode).
We assume that the \FourS decays to
pairs of \BpBm and \BzBzb mesons with equal probability and we do not include
any additional uncertainty due to this assumption.

We have identified several sources of systematic uncertainty as significant. The number of \BB pairs
in the data sample
is known with an uncertainty of 1.1\%. The uncertainties in the \Dz branching ratios are taken 
from Ref.~\cite{ref:pdg}. We determine 
the systematic errors arising from uncertainties in track,~\KS and \piz reconstruction and in kaon 
identification from studies of high statistics data control samples. 
The uncertainty in the track reconstruction efficiency is determined to be 0.8\% 
per track originating from the interaction region. There is an additional uncertainty of 3\% arising from
the knowledge of the \KS reconstruction efficiency. 
\begin{table}[tbh]
\caption{Systematic uncertainty estimates for each of the three \Dz decay samples.}
\center
\begin{tabular}{ l c c c }
\hline\hline 
Source    & \multicolumn{3}{c}{Uncertainty (\%)} \\
\hline
          & \KPi & \KPiPiz & \KThreePi\\
\hline
Number of \BB events          & 1.1 & 1.1 & 1.1 \\
Simulation statistics         & 5.6 & 6.5 & 6.0 \\
\Dz branching ratio           & 2.4 & 6.2 & 4.2 \\
Tracking efficiency           & 2.4 & 2.4 & 4.0 \\
\KS efficiency                & 3.0 & 3.0 & 3.0 \\
Particle identification       & 2.0 & 2.0 & 2.0 \\
\piz efficiency               &   -   & 5.0 &  -    \\
Peaking background            & 2.3 & 1.4 & 3.1 \\
\Kstarm lineshape             & 3.0 & 3.0 & 3.0 \\
Data/simulation differences   & 1.4 & 2.4 & 2.1  \\
\hline
Total                         & 8.6 & 11.9 & 10.3 \\
\hline\hline
\end{tabular}
\label{tab:syst}
\end{table}
The charged kaon identification leads to a systematic 
uncertainty of 2\%, and the \piz reconstruction to a systematic uncertainty of 5\%.
The systematic error from the
knowledge of the peaking background is taken 
from the studies of various simulated data samples described above.
An additional uncertainty from the knowledge of the \Kstarm lineshape has
been determined to be 3\%.
Finally, we include the errors on the correction factors determined from the \BmtoDzpim sample.
We have studied 
the uncertainty in the parameterization of the background and of the signal by repeating the
\mes fits with different combinations of 
parameters of the functional form fixed to values obtained either from simulation or from studies of
sideband regions in \DeltaE. We conclude that the systematic uncertainty from this source is negligible. 
A summary of the systematic errors estimate is shown in Table~\ref{tab:syst}.

The resulting \B branching fractions 
corresponding to three different \Dz decay modes
are listed in Table~\ref{tab:bf}. 
\begin{table}[h]
\caption{Measured branching fraction \BRBmtoDzKstarm. The first errors are 
statistical and the second systematic.}
\center
\begin{tabular}{ l c }
\hline\hline 
Decay Mode & $\BR(10^{-4})$ \\
\hline 
\DztoKPi & 5.8$\pm$1.0$\pm$0.5 \\
\DztoKPiPiz & 5.8$\pm$1.2$\pm$0.7 \\
\DztoKThreePi & 8.7$\pm$1.5$\pm$0.9 \\
\hline
Weighted Average & 6.3$\pm$0.7$\pm$0.5 \\
\hline\hline
\end{tabular}
\label{tab:bf}
\end{table}
We determine the
weighted average of the three measurements, $\BRBmtoDzKstarm = (6.3\pm0.7\pm0.5)\times10^{-4}$,
taking into account the correlations between the systematic uncertainties. 
The result of this analysis is in good agreement with a previous measurement
by CLEO, $\BRBmtoDzKstarm = (6.1\pm1.6\pm1.7)\times10^{-4}$~\cite{CLEO_BF}.

In summary, we have studied the decay \BmtoDzKstarm, where 
the \Dz was
detected through its decays to \KPi, \KPiPiz and \KThreePi and the
\Kstarm through its decay to \KS\pim. 
We have measured the branching fraction 
$\BRBmtoDzKstarm = (6.3\pm0.7\pm0.5)\times10^{-4}$.
This is in good agreement with the previous measurement of this branching
fraction, and
significantly improves on its precision. In the future, with larger
data samples, this decay will be studied with
the \Dz reconstructed in \CP\ eigenstates.
Eventually it is hoped that decays of the form $\B\to D^{(*)} K^{(*)}$ can provide
important constraints on the angle $\gamma$ of the  Unitarity Triangle.
 
% Input the pubboard acknowledgements file
We are grateful for the excellent luminosity and machine conditions
provided by our \pep2\ colleagues, 
and for the substantial dedicated effort from
the computing organizations that support \babar.
The collaborating institutions wish to thank 
SLAC for its support and kind hospitality. 
This work is supported by
DOE
and NSF (USA),
NSERC (Canada),
IHEP (China),
CEA and
CNRS-IN2P3
(France),
BMBF and DFG
(Germany),
INFN (Italy),
FOM (The Netherlands),
NFR (Norway),
MIST (Russia), and
PPARC (United Kingdom). 
Individuals have received support from the 
A.~P.~Sloan Foundation, 
Research Corporation,
and Alexander von Humboldt Foundation.


\begin{thebibliography}{99}

\bibitem{CKM}
N. Cabibbo, Phys. Rev. Lett. {\bf 10}, 531 (1963); 
M. Kobayashi and T. Maskawa, Prog. Theor. Phys. {\bf 49}, 652 (1973).

\bibitem{gamma}
M. Gronau and D. Wyler, Phys. Lett. B {\bf 265}, 172 (1991); 
I. Dunietz, Phys. Lett. B {\bf 270}, 75 (1991); 
D. Atwood, G. Eilam, M. Gronau and A.\ Soni, Phys. Lett. B {\bf 341}, 372 (1995); 
D. Atwood, I. Dunietz and A. Soni, Phys. Rev. Lett. {\bf 78}, 3257 (1997).

\bibitem{gammakstar}
J. H. Jang and P. Ko, Phys. Rev. D {\bf 58}, 111302 (1998); 
M. Gronau and J. L. Rosner, Phys. Lett. B {\bf 439}, 171 (1998).

\bibitem{chargeconj} Charge conjugate decays are implied throughout this paper.

\bibitem{CLEO_BF}  CLEO Collaboration, R. Mahapatra {et al.}, Phys. Rev. Lett. {\bf 88}, 101803 (2002).


% The NIM detector performance paper
\bibitem{ref:babar}
\babar\ Collaboration, B.Aubert {\em et al.}, Nucl. Instr. Meth. A {\bf 479}, 1 (2002).

\bibitem{ref:pdg}
Particle Data Group, K. Hagiwara {\em et al.},  Phys. Rev. D {\bf 66}, 010001 (2002) and 2003 off-year partial
 update for the 2004 edition available on the PDG WWW pages (http://pdg.lbl.gov).



\bibitem{DalitzWeight}
CLEO Collaboration, S.Kopp {\it et al.}, Phys. Rev. D {\bf 63}, 092001 (2001).

\bibitem{ref:FoxWol}
G. Fox and S. Wolfram, Phys. Rev. Lett. {\bf 41}, 1581 (1978).

\bibitem{babarfisher}
\babar\ Collaboration, B. Aubert {\em et al.}, Phys. Rev. Lett. {\bf 89}, 281802 (2002).

\bibitem{ARGUS}
ARGUS Collaboration, H. Albrecht {\em et al.}, Phys. Lett. B {\bf 185}, 218 (1987);
{\em ibid.} {\bf 241}, 278 (1990).

\end{thebibliography}
\end{document}